\documentclass[fleqn,twoside]{article}
\usepackage{espcrc2}

\usepackage{amssymb}
\usepackage{times}
\newcommand{\pubjournal}[6]{#1, #2 #3 (#5) #4.}
\newcommand{\aap}{A\&A}
\newcommand{\adv}{Adv.~Spa.~Res.}
\newcommand{\apj}{ApJ}

\newcommand{\mnras}{MNRAS}
\newcommand{\prc}{PRC}
\newcommand{\prd}{PRD}
\newcommand{\gray}{$\gamma$-ray}
\newcommand{\de}{DE}
\newcommand{\icrc}{ICRC}

\title{Understanding limitations in the determination
of the diffuse Galactic \gray\ emission}

\author{Igor V.\ Moskalenko\address[su]{Hansen Experimental Physics Laboratory,
    Stanford University, Stanford, CA 94305}\address[kipac]{Kavli 
    Institute for Particle Astrophysics and Cosmology,
    Stanford University, Stanford, CA 94309}\thanks{Supported in part 
    by NASA APRA grant},
Seth W.\ Digel\address{Stanford 
  Linear Accelerator Center, 
  2575 Sand Hill Rd, Menlo Park, CA 94025}\addressmark[kipac],
Troy A.\ Porter\address{Santa Cruz 
  Institute for Particle Physics,
  University of California, Santa Cruz, CA 95064}\thanks{Supported in part
  by the US Department of Energy},
Olaf Reimer\addressmark[su]\addressmark[kipac]
and
Andrew W.\ Strong\address{Max-Planck-Institut 
  f\"ur extraterrestrische Physik,
  Postfach 1312, D-85741 Garching, Germany}}

\begin{document}

\begin{abstract}

We discuss uncertainties and possible sources of errors associated
with the determination of the diffuse Galactic \gray\ emission using
the EGRET data.   Most of the issues will be relevant  also in the
GLAST era.  The focus here is on issues that impact evaluation of dark
matter annihilation signals against the diffuse \gray\ emission of the
Milky Way.

\vspace{1pc}
\end{abstract}




\maketitle

\section{INTRODUCTION}\label{intro}

Diffuse emission (\de) from the Milky Way dominates the \gray\ sky.
About 80\% of the high-energy luminosity of the Milky Way comes from
processes in the interstellar medium (ISM).   The \de\ traces
interactions of energetic particles, primarily protons and electrons,
in the ISM, thus delivering information about cosmic-ray (CR) spectra
in distant locations \cite{moskalenko_strong05}.   The DE may contain
signatures of exotic physics,  e.g., interactions of dark matter (see
\cite{moskalenko_strong05} for references).   Calculation of the \de\
requires first calculating the CR spectra throughout the entire Galaxy
\cite{SMR2004}.   The components of the \de\ model ($\pi^0$-decay,
inverse Compton -- IC, bremsstrahlung) are not independent and can not
be arbitrarily re-scaled  (e.g., as in \cite{deboer2005}).   The \de\
is the celestial foreground for the study of \gray\ point sources and
the extragalactic \de\ which may contain information about the early
universe.

A comparison between the EGRET data and the Galactic \de\ model used
by the EGRET team reveals an ``excess'' above 1 GeV of approximately a
factor of 2 in all sky directions \cite{SMR2004,hunter,SMR2000}.
Therefore, we need to look at the uncertainties associated with both
the data and the model to understand the origin of this excess.   In
this paper, we consider the sources of systematic uncertainties in the
EGRET calibration, data handling, and in models of the \de.

\section{SOURCES OF UNCERTAINTIES}
\subsection{EGRET data analysis}\label{egret}
EGRET was in orbit between April 1991 and June 2000, and operated for
much longer than its 2-year design lifetime.   Owing to ageing of the
spark chamber gas and the limited number of gas changes possible
during the mission, the best  data taking occurred during the early
years of operation.  Details of pre- and in-flight calibrations are
given in \cite{calibration,esposito}.

The data that we discuss here were taken during Cycles 1--4, the
``prime'' years of EGRET operation.   These are the data that were
used to produce the 3rd EGRET catalog of \gray\ point  sources
\cite{catalog}, and so for these data subtracting the (generally
variable)  point source contribution to the \de\ can be done
accurately and correctly using the catalog fluxes.   The overall
exposure, accumulated during the individually pointed observations, is
rather uneven \cite{esposito}.

The \de\ model   used for the source detections was derived on the
assumption that the CR proton and electron spectra do not change shape
across the Galaxy, and that the CR density is proportional to the
surface density of gas \cite{hunter}.  This leads to incorrect
predictions at intermediate and high latitudes.

The calibration of an instrument in orbit is a complicated matter.
Issues that cannot be controlled may affect the instrument
performance, for example  the launch vibrations and accelerations,
while radiation exposure and unknown factors occur during the whole
mission lifetime.  Several failures occurred during EGRET's
operational years: the spark chamber readout (switched to the
redundant side), the gas circulation pumps, and some PMTs of the
trigger telescopes.  This  complicates the analysis since the symmetry
of the instrumental response was lost, and thus the direct comparison
to the pre-flight calibration.

The efficiency of EGRET for registering reconstructable events from
\gray\ detections decreased with time as the spark chamber gas aged.
Several times during the mission the gas in the spark chamber was
replenished to improve the degraded efficiency. Efficiencies were
determined based on normalization to ``standard candles" (pulsars)
where possible, or by comparing to the \de\ predictions
\cite{esposito}.  The individual EGRET viewing periods were corrected
for the varying spark chamber performance under the assumption that
the energy dependence of the correction did not vary with time
\cite{esposito}.   This results in time-dependent systematic
uncertainties.

Several other uncertainties also must be considered \cite{esposito}:
contribution of unresolved point sources; fisheye effect corrections
for events under large viewing angles; contamination by Earth albedo
events \cite{Petry2005}; a direction-dependent PSF which results in
residuals in source-subtracted \de\ maps; several orbital reboosts
which changed the particle background of the detectors; plus, late in
the mission, operation in different modes (narrow, strip) for which
the instrument performance is less well understood.

After these considerations, any model of \de\ must be convolved with
the instrumental PSF before comparison with the data. The effect is
large below 1 GeV affecting the overall spectral shape. This is
especially important for small areas on the sky. For comparisons
with data obtained by a different instrument, a deconvolution of the
data is preferred \cite{SMR2004}: in this case the procedure  is
model-dependent since the predicted spectrum is convolved with the PSF
first, and subsequently the appropriate factors are applied to the
data.

The EGRET team made corrections for the described effects taking into
account the limited statistics.   The quoted error bars
\cite{esposito} are not separable into statistical and systematic
parts (e.g., as in \cite{deboer2005}) because of the implicit time
dependence and interplay of the applied corrections.

\subsection{CR propagation}\label{cr_propagation}

The Galactic diffuse \gray\ emission is the product of particle
interactions with gas in the ISM  and the interstellar radiation field
(ISRF).   Therefore, its spectrum depends on the particle spectra
throughout the entire Galaxy \cite{moskalenko_strong05}.   Diffusion,
energy losses, and other processes change the particle spectra during
propagation in the Galaxy over long time periods.   Additionally, the
CR isotopic composition changes.   During propagation within the
heliosphere, solar modulation further alters the particle spectra.
These spectra are what can be measured directly using satellite- and
balloon-borne instruments.   To calculate the \gray\ emission we thus
need to ``reconstruct" the particle spectra in the ISM.

Gamma rays from $\pi^0$-decay are produced in the same processes as
other secondaries such as antiprotons and positrons, which can also be
used to test the models of the \de\ \cite{SMR2004,M1998}.  A realistic
model of CR propagation has to include the gas  and source
distributions, ISRF, nuclear and particle cross  sections and nuclear
reaction network, \gray\ production processes,  energy losses, and
solve the transport equations for the individual CR species.   The
propagation parameters, the diffusion coefficient, halo size,
Alfv\'{e}n speed, convection velocity, and so on, which depend on the
assumed propagation model, are derived by comparing model predictions
with CR data, such as secondary/primary nuclei ratios, e.g. B/C, and
radioactive isotope ratios like $^{10}$Be/$^{9}$Be.   Therefore, the
accuracy of the nuclear cross sections is one of the major concerns
for uncertainties in all subsequent analysis tasks.

Semi-empirical estimates of the nuclear cross sections
\cite{W-code,tsao99} have been tuned to match the isotopic production
cross sections between $\sim$400 and 700 MeV/n.   They can not provide
the same accuracy over the whole energy range which is required for CR
propagation calculations.   Furthermore,  the cross sections often
have resonances at energies below $\sim$200 MeV/n.   These are
particularly important for propagation models including diffusive
reacceleration.   This adds to the overall uncertainty in
determination of the propagation parameters (e.g., see \cite{MMS01}).

The interpretation of the peak in the B/C ratio depends on the assumed
propagation model.  Different models predict different rigidity
dependencies for the diffusion coefficient.  Therefore, determination
of the diffusion parameters is model-dependent.   Two examples  are
the reacceleration \cite{seo} and convection \cite{Zirakashvili96}
models.   Reacceleration is a Fermi 2nd order acceleration in the ISM;
it predicts a diffusion coefficient rigidity power-law index of 1/3.
Convection is essentially a Galactic wind moving away from the
Galactic plane which affects mostly particles below a few GeV;
consistency with high-energy data requires a diffusion coefficient
rigidity index of 0.6.   A recently developed ``damping'' model
includes CR particle -- MHD wave interactions which lead to a concave
shape in the diffusion coefficient, where the true rigidity dependence
is derived  via self-consistent modeling \cite{ptuskin06}.

Above 1 GeV, $\pi^0$-decay and IC scattering are the dominant \gray\
production processes. The $\pi^0$-production cross section is based
on data from the 1960s that themselves have large error bars.   The
proposed parameterizations (\cite{Dermer86b} and references therein)
fit these data well, but they do not provide the required accuracy.
New details are still being added, e.g.\ the diffraction dissociation
and violation of the Feynman scaling \cite{kamae2005}, while various
Monte Carlo event generators produce results which differ by
$\sim$20--30\% \cite{kelner}.   The IC scattering contribution is
dependent on the underlying model of the ISRF, and the  treatment of
the scattering process using the proper ISRF angular distribution
\cite{MS2000}.   Conventionally, the ISRF is assumed to be isotropic,
which is only true for the cosmic microwave background.   Using the
full angular distribution for the ISRF will alter the model
predictions \cite{MS2000}.

The random nature of CR sources leads to fluctuations of CR intensity
in space and time. Very high energy (VHE) electrons cannot propagate
far from their sources due to rapid energy losses. Their interstellar
spectrum is expected to exhibit imprints from their nearby sources
\cite{kobayashi03}.

Finally, ``solar modulation'' changes the interstellar spectra of CR
particles below $\sim$20 GeV/n during their propagation within the
heliosphere. It is a combination of the effects of convection by the
solar wind, diffusion, adiabatic cooling, drifts, and diffusive
acceleration. The theory of solar modulation is far from complete
\cite{fichtner05}; current  models are based on the solution of
Parker's transport equation. The unknown is the interplay of the
different terms and the CR spectrum in the ISM.  Spherically symmetric
solutions, the force-field \cite{force-field}, and Fisk \cite{fisk}
approximations are most often used. However, they include only the
effect of adiabatic losses. The Pioneer, Voyager\footnote{The Voyager
1 spacecraft  is currently approaching the outer boundary of the solar
system \cite{stone05}.}, and Ulysses missions contributed  to
understanding the global aspects of modulation.

\subsection{Astrophysical input}\label{astro}

Most of the interstellar gas is in the form of neutral hydrogen
H~{\sc i} and H$_2$.    The H$_2$ gas is not observed directly but
via the mm-wave spectral line of a rotational transition of CO.   A
proportionality is assumed between CO surface brightness and H$_2$
column density.   H~{\sc i} gas is detected by its  21 cm line.
Determinations of H~{\sc i} column densities rely on
difficult-to-verify assumptions about the spin temperature of the gas,
and in many directions colder H~{\sc i} gas is seen in absorption
against warmer background gas, making overall distributions of  H~{\sc
i} difficult to disentangle.   The spatial distribution of the gas is
derived using velocity measurements and assuming a rotation curve for
the Milky Way.   However,  streaming motions and velocity dispersions
between individual gas clouds introduce systematic uncertainties.   In
the outer Galaxy the gradient of velocity approaches zero at large
distances.   In the inner Galaxy the ``kinematic distance'' is double
valued  except toward the Galactic center and anticenter, where the
line-of-sight velocities provide no useful kinematic information.
Therefore, no unique solution exists for the 3D distribution of gas.
Studies starting with the same data produced gas distributions which
are different in many details \cite{hunter,pohl}.

The CR source distribution is also not very well known.  The pulsar
distribution \cite{lorimer} is too peaked to reproduce the  CR
gradient in the Galaxy. Consistency with the latter requires the
$X$-factors (essentially H$_2$/CO ratio) to increase  with the
distance from the Galactic center \cite{S2004}.   Such behavior indeed
is observed in other galaxies \cite{israel01}.

The ISRF modeling incorporates a stellar distribution model, a model
for the dust distribution and properties, and a treatment of
scattering, absorption, and subsequent re-emission of the stellar
light by the dust \cite{PS05}.  The stellar emission model
incorporates details for the distribution of  stellar types in
discrete geometrical components (bulge, disc, spiral arms, etc.),
adjusted to agree with results from experiments such as 2MASS, SDSS,
and others.   Dust is modeled with a mixture of PAH, graphite, and
silicate.   Details of the dust absorption and scattering
efficiencies, abundances, and size distribution are included in the
scattering and heating calculations.   The dust is assumed to follow
the Galactic gas distribution and  metallicity gradient.
Uncertainties in each of these inputs contribute to the  overall
uncertainty in the ISRF modeling.   The model can be compared with the
ISRF only at the solar system position.  The deviations from an
earlier model \cite{SMR2000,MPS2006} give an idea of possible
systematic errors.

\section{DISCUSSION AND CONCLUSION}\label{conclusion}

We discussed only the most obvious sources of uncertainties; the
systematic effects are numerous, but not all of them are equally
important.  Considering the many sources of uncertainties, the
agreement between the predictions of conventional \de\ models with the
EGRET data is remarkable: the discrepancy is only a factor of two at
worst; this is the famous ``GeV excess.''   With some reasonable
assumptions, such as CR intensity fluctuations, agreement with the
data can be obtained in the framework of conventional astrophysics
\cite{SMR2004}.

The Gamma Ray Large Area Space Telescope (GLAST), to be launched in
2007,  will have improved sensitivity, angular resolution, a wider
field of view, and uses a different technology \cite{McEnery2004} than
EGRET.  Therefore, many of the instrumental uncertainties associated
with EGRET will become irrelevant.    However, issues associated with
CR propagation, gas distribution, and other astrophysical input will
remain waiting for the next generation of instruments.

\end{document}